\documentstyle[multicol,aps,epsf]{revtex} 
\begin{document}

\title{Decoupling and melting in a layered superconductor}

\draft
\author{N.K. Wilkin \cite{add} and Henrik Jeldtoft Jensen}

\address{Department of Mathematics,\\ Imperial College, 180 Queen's Gate,\\
London, SW7 2BZ, United Kingdom}

\maketitle
\begin{abstract}

We report results for a 3D simulation of a layered
superconductor. There are two significant temperatures. The first
corresponds to melting of the vortex lattice in the $a$-$b$ plane
($T_{\rm ab}$) and the second to decoupling of the layers ($T_{\rm
dc}$) with $T_{\rm ab} \le T_{\rm dc}$. The decoupling is found to be
a first order transition with an associated entropy of $\sim
0.25-0.4{\rm k}_B$/pancake. The melting has no obvious thermodynamic
signature, and could be a crossover.  The width of the intermediate
regime $T_{\rm ab} <T<T_{\rm dc}$, decreases with increasing anisotropy
such that for the more anisotropic system we cannot distinguish the
melting and decoupling temperatures.

\end{abstract}

\begin{multicols}{2}

The phase diagram for high temperature superconductors is known to be
complicated by the enhanced thermal fluctuations and the large
anisotropy due to the layered structure of the crystals. Most
measurements to date have been transport and thus indirect, making it
difficult to ascertain the order of phase transitions and whether they
are artifacts of the measuring current.  However, Zeldov {\em et al.}\
\cite{Zeldov_Nat_95} have recently presented direct thermodynamic
evidence of a first order transition in Bi-2212 using local
magnetization techniques. Subsequently first order transitions have
also been reported in the magnetization data of YBaCuO$_{7-\delta}$ by
Liang {\em et al.}\ \cite{Liang_96} and Welp {\em et al.}\
\cite{Welp_96} and in the specific heat data of Schilling {\em et al.}
\cite{Schilling_96}. The anisotropic nature of these
materials is such that we could expect to lose the groundstate at
different temperatures in the $a$-$b$ plane (corresponding to the
Cu-O layers) and the $c$-axis along which the external magnetic field is
applied. If we assume a vortex lattice at low temperatures (where it
is common to think of a vortex in these layered systems as pancakes in
the $a$-$b$ planes connected via Josephson vortices) then the
transition corresponding to loss of order in the `$a$-$b$' plane is an
`$a$-$b$ plane melting' transition. The loss of order only along the
c-axis is a `decoupling' transition.  The predicted field-temperature
dependences are similar for these two transitions such that although
the Zeldov {\em et al.}\ data fits the decoupling scenario marginally
better it is not decisive. It is of course possible that the two
transitions occur simultaneously.

In this letter we investigate the $a$-$b$ melting and decoupling
transitions within a layered pancake model.  We study the transition
as a function of fixed magnetic field, B (density of vortices), which
is always perpendicular to the layers (along the c-axis), and vary the
temperature. We can also alter the effective anisotropy of our system
by changing the strength of the coupling between the layers. In order
to be able to study the loss of the vortex lattice order we let the
vortex position vary continuously. An underlying discrete lattice may
lead to spurious phases\cite{Franz_94}. An essential difference from
previous simulations \cite{Ryu_96} is the manner in which we allow the
correlation along the c-axis to vary with temperature by allowing the
vortices to cross, or in `pancake' language to interchange neighbours
in adjacent planes, when it is energetically favorable. It is this
decoupling of the layers which leads to an entropy jump of magnitude
comparable to that seen experimentally. It is important to note that
the simulation {\em does not} include free vortex loops in the $a-b$
plane. It has been suggested that these loops are a requirement
\cite{NSH96} for seeing the large entropy jumps that are measured
experimentally.

We find that we can lose the correlation in the $a$-$b$ plane and
along the c-axis at different temperatures, with the width of the
intermediate regime determined by the anisotropy of the system.  In
the more anisotropic case melting and decoupling occur over a
temperature range smaller than our resolution. However, in the less
anisotropic case the $a-b$ order is lost at a lower temperature than
the c-axis order. There is a large entropy jump which occurs
concurrently with the loss of order in the c-direction,
$0.2-0.4k_B$/pancake. There is no obvious thermodynamic signal
associated with the loss of order in the $a-b$ direction, such that it
may be a crossover.  In terms of the pancakes this means that by
raising the temperature the system changes from a 3D vortex-line
lattice to a 3D line-liquid to a 2D pancake-liquid or gas, with the
existence of the intermediate regime being dependent on the
anisotropy.


To equilibriate our system we use Langevin dynamics and periodic
boundary conditions in all directions with the temperature introduced
via a noise term \cite{Brass_89}. We neglect the additional
temperature dependencies of the penetration depth and other length
scales. We focus purely on the vortex lattice aspect of the melting --
that is only vortex loops representing fluctuations in the positions
of the flux lines are included.  We believe that our model includes
the essential symmetry and topology of the vortex system. The Gaussian
form of the potentials are chosen for numerical viability. Previous
work in 2D has demonstrated that the qualitative behaviour is
unchanged between the Gaussian and realistic potentials
\cite{Jensen_88,Koshelev_92}. The pancakes in the planes have an
in-plane repulsive interaction between them, in our case modeled by a
Gaussian potential $U_{{\rm v}{\rm v}'}^{ll}=A_{{\rm
v}}\exp(-r^2/\xi_{\rm v}^2)$, where $r$ is the in-plane distance
between the vortices, $ \xi_{{\rm v}}$ is the in-plane vortex range,
and $A_{\rm v}=1$ is the (fixed) strength of the vortex potential.
 
Across the layers the interaction is more complicated. In order to
allow the pancakes to change neighbours freely throughout the
simulation we cannot define unbreakable `strings' of pancakes. The
`strings' effectively prohibit vortex cutting and re-connecting a
mechanism which is thought to be important in the vicinity of the
melting temperature; and is needed for the loss of long range phase
coherence as seen in the pseudo-transformer experiments
\cite{delacruz_94}. It has been shown by Clem \cite{Clem_91} that in
order to model the electromagnetic interactions across the layers only
pair-wise potentials are needed. However, Bulaevskii {\em et al.}\
\cite{Bulaevskii_92a} have shown that including the lowest order terms
of the Josephson coupling makes three and four-body terms equally
necessary. Hence we include both an attractive two body potential
$U_{{\rm v} {\rm v}'}^{l l'}(r_i^l,r_j^{l'})= -A_{\rm
l}\exp(-(r_i^l-r_j^{l'})^2/\xi_{l^2}) $ (all $r$ are in-plane) and a
repulsive 3-body potential which stabilizes the triangular lattice
phase.
\begin{equation} 
U_{{\rm v} {\rm v}'{\rm v}''}^{l l l'}(r_i^l,r_j^l,r_k^{l'})= A_{{\rm 3b}} 
e^ {- \left((r_i^l-r_j^l)^2+(r_j^l-r_k^{l'})^2
+(r_k^{l'}-r_i^l)^2\right)/\xi_{\rm 3b}^2}.
\end{equation}

$A_{{\rm 3b}}$ and $\xi_{\rm 3b}$ are the amplitude and range of the three body
potential. It acts by excluding three or more pancakes (two in one
layer and a third in an adjacent layer) from finding their equilibrium
location to be within a coherence length in the x-y plane.

 We change the anisotropy of the system by varying the
interlayer coupling parameter $A_l$, thus we have used three sets of
parameters in our simulations, $A_l=0.05$, $A_l=0.2$ and $A_l=0.5$. The other
parameters are: $A_{\rm v}=1$ and $A_{\rm 3b}=A_l$, for the amplitudes of the
potentials and $\xi_{{\rm v}}=0.6$, $\xi_{l}=0.3$, $\xi_{{\rm 3b}}=\sqrt{2}
\xi_{l}$ for the ranges in the Gaussians.

In order to detect phase changes we measure a number of physical
properties of the system. The most pronounced feature is seen in the
specific heat/pancake which has a peak centred on $T_{\rm dc}$. The
origin of the peak is identified to be a first order transition by
applying the Lee-Kosterlitz energy binning technique \cite{LeeKost90}.

The shear modulus, $c_{66}$ has recently been measured experimentally
\cite{Kwok_96} and can also be estimated numerically. It should vanish
when the system becomes liquid in the $a$-$b$ planes. It is measured
by applying a shear wave across the system and then holding the system
pinned in two channels whilst allowing the rest if the system to relax
\cite{Jensen_89}. We obtain $c_{66}$ via, \begin{equation} \Delta
U=\frac{1}{2}\mbox{ } c_{66}\int {\rm d}x \int {\rm d} y
\left(\frac{\partial u}{\partial x}\right)^2.
\end{equation} There are fluctuations in the data due to the finite
temperature but the drop to zero of $c_{66}$ is well defined.

Using the same technique we also measure the Josephson contribution to
the tilt modulus, $c_{44}^{\rm J}$, which corresponds to shearing
across the layers. Within our model there is no contribution of the
magnetic field energy so that $c_{44}^{\rm J}$ should be finite for a
line liquid but will be zero for a 2d pancake liquid.

That the system is becoming thermally excited is readily seen by the
onset of self-diffusion. The time dependence of the displacements
$r^2(t)$ of the pancakes indicates the nature of the
diffusion. Independent pancakes or rigid rods of pancake stacks behave
as $ r^2 \propto t $. Whereas diffusion of flexible unbreakable lines
follow $r^2 \propto t^{1/2}$.

In the clean system that we are studying an indicator of the in-plane
melting is given by monitoring via a Voronoi construction the
co-ordination number of the vortices which is six for the triangular
lattice groundstate. We find that the appearance of vortices with
coordination number other than six and the onset of diffusion match to
the accuracy that we have sampled the temperature.  Finally the
structure factor is used to confirm the nature of the vortex phases
present in the simulation.
 

We shall first consider the results for the larger interlayer coupling
$A_l=0.5$. We started from the groundstate Abrikosov lattice at T=0
with a system size chosen such that it would be commensurate with the
lattice. We then slowly heated the system up and observed the physical
quantities detailed above. We ran for system sizes with either 8x8 or
12x12 vortices in the $a$-$b$ plane and up to 64 layers. At
approximately $T_{\rm ab}=0.084$ we found that the diffusion
coefficient `D' abruptly became non-zero, see Fig.~\ref{fig:onset}. We
then continued to heat the system and studied the exponent $\alpha$ of
the diffusion $R^2\sim t^{\alpha}$. For all systems it was clear that
the exponent was initially in the range $1/2 <\alpha <1$. For the
system size 12x12x16 we found that $\alpha$ linearly interpolates
between $\alpha=1/2$ at $T=T_{ab}$ and $\alpha=1$ at a higher
temperature $T_{dc}$. See Fig.~\ref{fig:onset}. That is from the
diffusion alone we have evidence of a liquid phase in which there are
lines present although their coherence decreases (susceptibility to
breaking increases) with temperature up to $T_{dc}$, at which point
the entire system becomes a 2D pancake liquid.

Using a Voronoi construction we measure the onset of miscordination of
the vortices. We find that the appearance of vortices with other than
six-fold symmetry coincides with the onset of diffusion, $T_{\rm ab}$
and the temperature at which this saturates coincides with that at
which the diffusion exponent $\alpha$ becomes equal to one, $T_{\rm
dc}$ see Fig.~\ref{fig:onset}. The melting in the $a-b$ plane could be
a KTHNY transition \cite{KTHNY}. This requires a phase in which paired
5 and 7 fold coordinated vortices are present. However, the 5 and 7
fold vortices occur abruptly at $T_{\rm ab}$ and there is no sign of
pairing at any temperature.

The specific heat has a peak centered around $T_{dc}$ whose height
increases with increasing system size, see Fig.~2. In order to calculate the
entropy from the Lee-Kosterlitz procedure it is necessary to have a
sufficiently large system such that $L<<\xi$,this occurs for a system
size 8x8x32. The resulting entropy is $\sim 0.25k_B$/pancake.

In agreement with the above scenario the shear modulus, $c_{66}$ which
measures the rigidity in the $a$-$b$ plane falls to zero in the
vicinity of the lower temperature $T_{ab}$ and the tilt modulus,
$c_{44}$ falls to zero around the temperature $T_{dc}$. It is expected
that $c_{44}^{\rm J}$ should be non-zero in a vortex line liquid.
These results hold true for all system sizes\cite{unpub1}. This is in
agreement with losing the $a$-$b$ plane order at
$T_{ab}$ but still retaining some $c$-axis correlation up to $T_{dc}$.

We can also see that the c-axis coherence persists to the higher
temperatures $T_{dc}$ by measuring the relative diffusion of the
pancakes with respect to their neighbours in adjacent layers, the
diffusive width. We find that the width of the pancake stacks becomes
larger than $a_0/\sqrt{2}$ above $T_{\rm dc}$ corresponding to
disintegration of the stacks, see Fig.~\ref{fig:wiggle}.

This scenario of a 3D vortex lattice melting to a 3D line liquid and
then subsequently to a 2D pancake liquid is borne out by the structure
factor, see Fig.~\ref{fig:struct}.  At temperatures below $T_{\rm ab}$
we find rings of well defined spots in k-space consistent with a real
space triangular lattice.  For temperatures in the range $T_{\rm ab}
<T < T_{\rm dc}$ we see a ring of intensity consistent with a liquid
structure. The average height of the amplitude of the ring is a
measure of the correlations along the c-axis. Finally, as the
temperature becomes larger then $T_{\rm dc}$ the amplitude of the the
ring decreases rapidly signaling that the stacks of pancakes evaporate
and therefore the correlations along the c-axis are lost.

In order to see how the transitions changed when we increased the
anisotropy (made the system more Bi-2212 like) the study was repeated
with $A_l=0.2$. The temperature scale at which the system started to
break up was a little lower, around $T \simeq 0.062$. In this case the
temperatures for melting and decoupling cannot be differentiated.
Hence,there is no apparent intermediate regime where the diffusion is
line-liquid like, the miscoordination reaches saturation over a very
small temperature range and $c_{66}$ and $c_{44}^{\rm J}$ seem to die
off at approximately the same temperature. The peak in the specific
heat is very sharp -with the small fluctuations in the temperature
leading to large error bars on the peak height, hence direct
evaluation of the entropy is inaccurate. However, the Lee-Kosterlitz
energy plots yield an entropy of $\sim 0.4 k_{\rm B} $/pancake for all
system sizes greater than 8x8x8.  If this is the regime into which
Bi-2212 falls it would appear that the system sublimates to a pancake
gas \cite{Fuchs97}.

In this layered 3D simulation we include vortex crossing and allow the
pancakes to move continuously in the $a$-$b$ plane.  We find two
significant temperatures for the loss of 3D lattice order.  T$_{\rm
ab}$, at which the system melts in the $a$-$b$ plane and a second
temperature $T_{\rm dc}$ where the layers decouple.  For the more
anisotropic case the two temperatures coincide and we observe a
 first order
transition with an entropy jump of the same order of magnitude as seen
experimentally away from $T_c$. By making the system less anisotropic
the two temperatures separated, with the entropy and first order
nature remaining associated with the decoupling transition. It is
important to note that these results have been obtained in the absence
of ab-plane vortex loops, which have hitherto been claimed to be
essential in producing the entropy jumps seen experimentally. We
believe that our pancake model represents faithfully the physics of
the vortex array degrees of freedom. \cite{Nelson88}.

We are grateful for interesting discussions with Mike Moore, Ted
Forgan; John Clem for refering us to Ref.~\onlinecite{Bulaevskii_92a},
and Arrigo Triulzi for technical assistance. N.K.W thanks the School
of Physics, Birmingham University for hospitality whilst some of this
work was carried out. N.K.W.\ and H.J.J.\ were supported by the EPSRC
grant no.\ Gr/J 36952.

\begin{figure}
\narrowtext
\caption{The diffusion exponent $\alpha$ after the onset of diffusion and the
fraction of five fold co-ordinated pancakes as a function of
temperature. The inset show the onset of diffusion for the two
different anisotropies studied. The more isotropic system starts to diffuse
at a higher temperature.}
\label{fig:onset}
\end{figure}

\begin{figure}[t]
\narrowtext
\caption{Specific heat peaks (per pancake) for different system sizes with 
interlayer coupling $A_l=0.2$.The figure shows the specific
heat for two system sizes. For the 8x8x32 the width of the peak has
become narrower than our temperature resolution and the height of the
peak has a large error bar, $\pm20$ associated with it. The insert
shows the Lee-Kosterlitz binning for T=0.063 for the 8x8x16 system.
From this we estimate the entropy associated with the transition.
}
\label{fig:spec_Al0.2}
\end{figure}

\begin{figure}
\narrowtext
\caption{Disappearance of the  correlation along the c-axis as we
increase the temperature for $A_l=0.5$, $12\times 12 \times16$.}
\label{fig:wiggle}
\end{figure}

\begin{figure}
\narrowtext
\caption{The structure factor for the 8x8x8 $A_l=0.5$ system. The top
figure is for $T<T_{\rm ab}$, the middle for $T_{\rm ab} <T <T_{\rm
dc}$ and the bottom for $T>T_{\rm dc}$. Figure available at http://th.ph.bham.ac.uk/nkw/res/sup1.ps}
\label{fig:struct}
\end{figure}

\end{multicols}
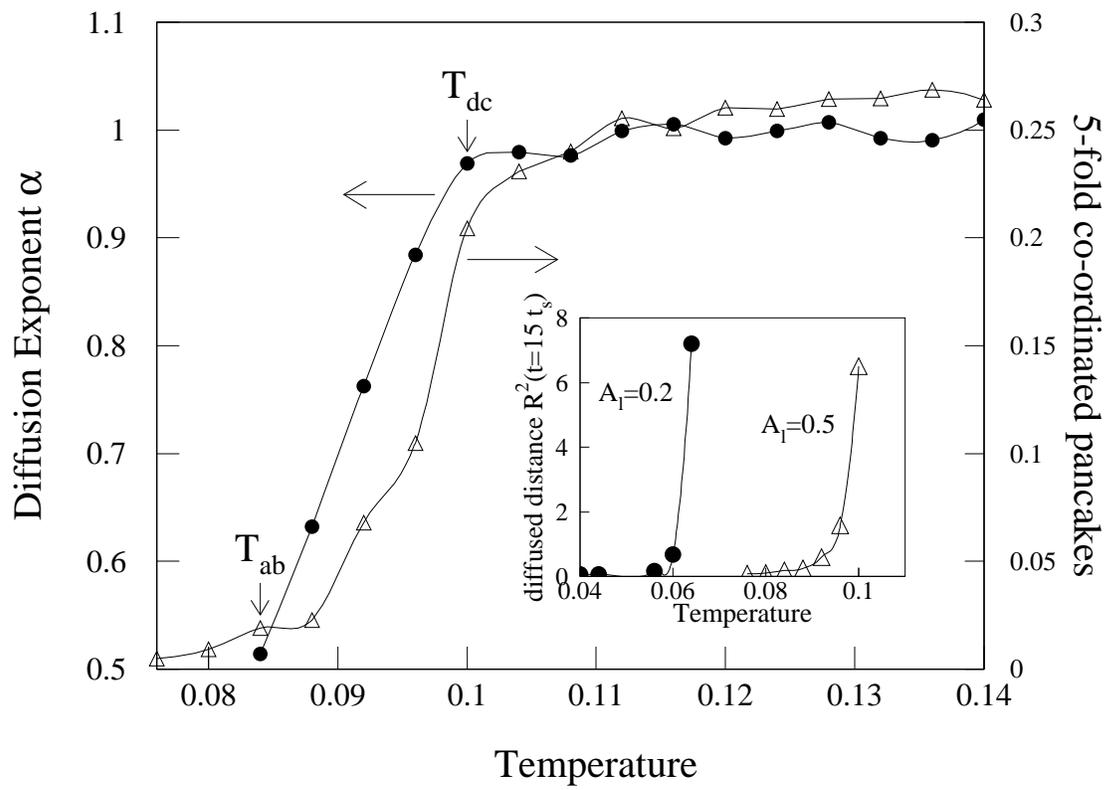
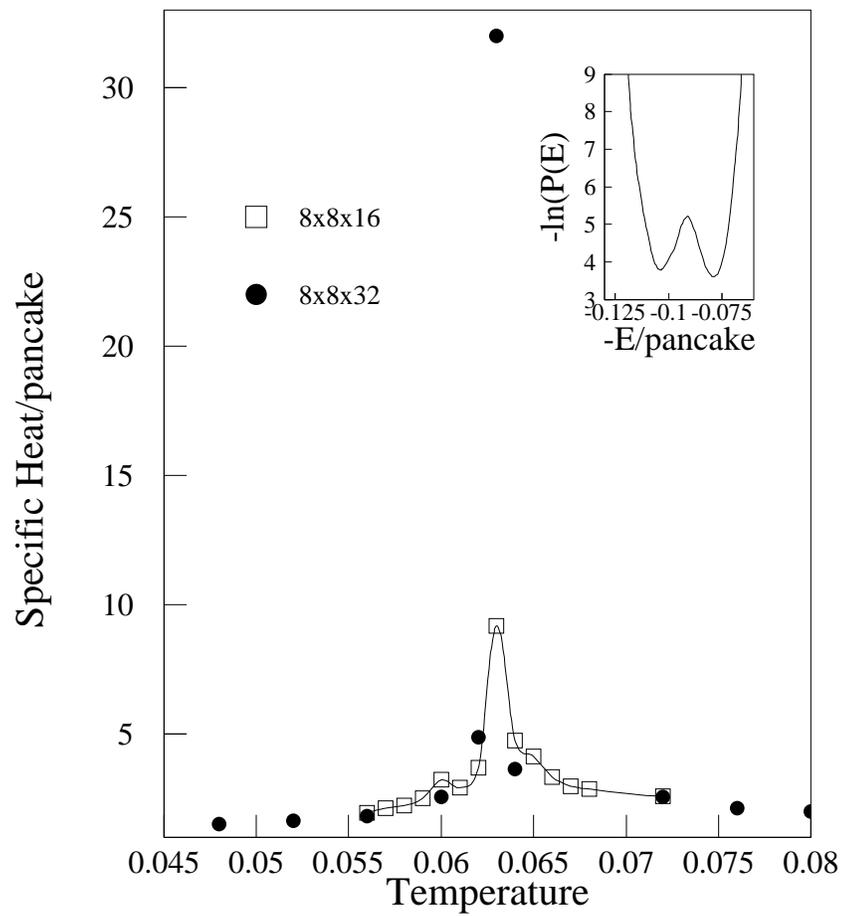
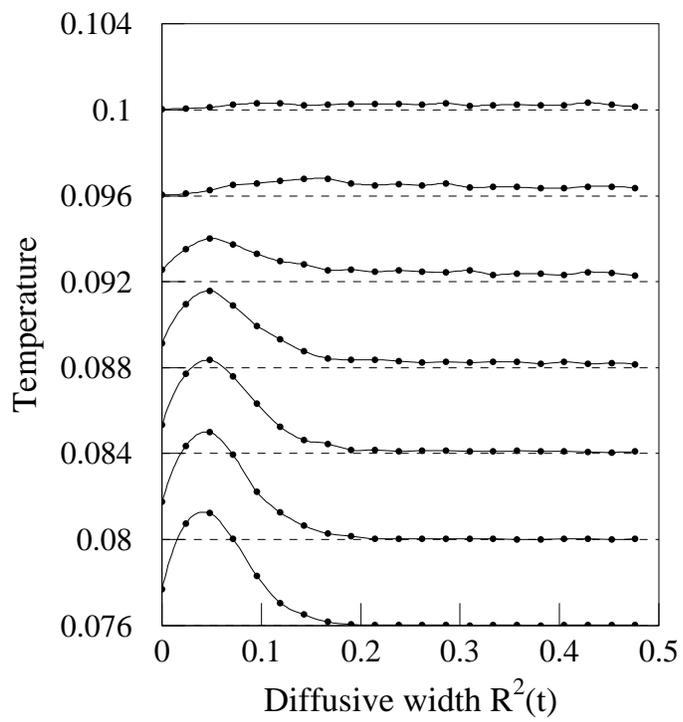
\end{document}